# Switching current distributions in InAs nanowire Josephson junctions

Bum-Kyu Kim[1] and Yong-Joo Doh[1,2*]

[1]Department of Applied Physics, Korea University Sejong Campus, Sejong, 339-700, Korea

[2]Department of Physics and Photon Science, School of Physics and Chemistry, Gwangju Institute of Science and Technology(GIST), Gwangju, 61005, Korea

[*]Corresponding author: yjdoh@gist.ac.kr

**Abstract**

We report on the switching current distributions in nano-hybrid Josephson junctions made of InAs semiconductor nanowires. Temperature dependence of the switching current distribution can be understood by motion of Josephson phase particle escaping from a tilted washboard potential, fitted well to the macroscopic quantum tunneling, thermal activation and phase diffusion models, depending on temperature. Application of gate voltage to tune the Josephson coupling strength enables us to adjust the effective temperature for the escaping process, which would be promising for developing gate-tunable superconducting phase qubits.

A Josephson junction[1] (JJ) or a superconducting weak link is made by sandwiching a thin layer of a nonsuperconducting material between two superconducting electrodes. Since a JJ can be regarded as an artificial atom of superconducting phase particles, whose confinement energy is determined by the Josephson coupling strength, it constitutes a basic building block of superconducting quantum bit (qubit).[2] When the spacing layer of the JJ is made of a semiconducting material, it is possible to form a gate-tunable JJ, where the Josephson coupling strength and the resulting Josephson effects can be tuned by the application of the gate voltage.[3] Recent advance in nanotechnology has led to a major breakthrough in the fabrication and performance of nano-hybrid gate-tunable JJs based on a semiconductor nanowire,[4] carbon nanotube,[5] graphene,[6] and topological insulator.[7] Furthermore, the nano-hybrid JJs provide a useful platform to observe Majorana fermions[8,9] and to develop a gate-tunable transmon qubit,[10] [11] which would be promising for realizing a protectable and scalable qubit.

The supercurrent switching event in a current-biased Josephson junction is equivalent to an escaping motion of a Josephson phase particle in a tilted washboard potential[12] determined by the Josephson potential and the bias current, which is given by $U(\varphi) = -E_{J0}[\cos(\varphi)+(I/I_{C0})]$. Here, $\varphi$ is the phase difference across the junction, $I_{C0}$ is the fluctuation-free switching current, and $E_{J0} = \hbar I_{C0}/2e$ is the Josephson coupling strength. It is well known that the phase particle can escape from the potential well via macroscopic quantum tunneling [13,14] (MQT), thermal activation[15] (TA), or phase diffusion[16,17] (PD) processes, depending on temperature and Josephson coupling energy.

In this paper, we report on the stochastic switching current distribution in InAs nanowire (NW) JJ. The switching current distributions in the NW JJ were obtained with varying temperature and gate voltage. The escape rate of a phase particle was numerically obtained from the switching current distribution data and fitted well by an appropriate theoretical model. The effective escape temperature $T_{esc}$ of phase particle was also obtained by fitting the switching current distribution to the escape model, confirming the validity of the fitting results. Application of gate voltage induces change of $T_{esc}$, indicating a gate-tunable $T_{esc}$. We believe that our observations provide useful information for developing nano-hybrid phase qubit based on semiconducting nanowires.

InAs NWs were grown via a catalytic process by using vapor-liquid-solid mechanism[4]. After the NWs were transferred to a highly $p$-doped silicon substrate covered with 250 nm-thick oxide layer, e-beam evaporation of Ti(10 nm)/Al(120 nm) and e-beam lithography were followed to form metallic electrodes. Before metal deposition, the NW surface was deoxidized using buffered hydrofluoric acid for 6 s to ensure transparent contacts. Scanning electron microscope (SEM) image of a typical NW device is shown in the inset of Figure 1. The channel length and diameter of the NW are $L$ = 250 (150) nm and $\phi$ = 100 (70) nm for device D1 (D2), respectively.

Current-voltage ($I$-$V$) curves of InAs NW JJ at different temperatures are displayed in Fig. 1a. It is clearly shown that the $I$-$V$ curves exhibit a hysteretic behavior with a critical current $I_C$, switching from superconducting state to resistive one, and a return current $I_R$ (vice versa), while the hysteresis disappears at temperatures higher than $T$ = 0.5 K. The hysteretic $I$-$V$ curve is attributed to a shunted capacitor formed between source and drain electrodes via the conductive Si substrate[4], Joule heating [18], or an effective capacitance due to a diffusive motion of conduction electrons in the nanostructures[19,20]. Since the dissipation power is very low ($P_{Joule}$ ~ 4 pW), the self-heating effect is ruled out to explain the hysteresis. The effective capacitance is given by $C_{eff} = \hbar/R_N E_{Th}$, where $\hbar$ is the Planck's constant divided by $2\pi$, $R_N$ (~ 330 $\Omega$) is the normal-state resistance of the junction and $E_{Th}$ is the Thouless energy. Since $E_{Th} \sim 0.1$ meV (will be discussed below), it is estimated to be $C_{eff} \sim 20$ fF. We note that this capacitance value is comparable to the one estimated from the switching current distributions, as will be discussed later.

Temperature dependences of $I_C$ and $I_R$ are displayed in Fig. 1b. As we increase temperature, $I_C$ is exponentially decreased near the critical temperature $T_C = 1.1$ K. This $I_C(T)$ behavior fits well the diffusive JJ model in a long junction regime[21], $eI_CR_N = aE_{Th}[1-b\exp(-aE_{Th}/3.2k_BT)]$, where $a$ and $b$ are fitting parameters and $k_B$ is the Boltzmann constant. From the normal-state transport, $E_{Th} = \hbar D/L^2$ is obtained to be $\sim 0.1$ meV, where $D = v_Fl_e/3 \sim 95$ cm$^2$/s is the diffusion constant, $v_F \sim 10^8$ cm/s is the Fermi velocity and $l_e \sim 30$ nm is the elastic mean free path of InAs NW, respectively.[4] The best fit result (solid line in Fig. 1b) is obtained with $a = 0.54$ and $b = 1.18$.

Temperature dependent behavior of the switching current ($I_C$) distribution is displayed in Fig. 2a, where the switching events were recorded 2,000 times with a voltage criterion of $V_{th} = 8$ μV at each temperature. We used triangle-wave current with a sweep rate d$I$/d$t$ = 32 μA/s for the switching measurements. It is evidently shown that the $I_C$ distribution is very sharp at higher temperatures and becomes wider at intermediate temperatures and turns into a sharp one again at very low temperatures. Temperature dependence of the normalized standard deviation (SD), $I_{SD}/I_{C0}$, in Fig. 2b reveals three distinct switching regimes of MQT, TA, and PD, which resembles the graphene-based JJ.[22,23] Below $T = 0.1$ K, the normalized SD is nearly insensitive to temperature, indicating that the escape rate of the phase particle over the potential barrier is governed by MQT process[24]. For $0.1$ K $< T <$ $0.2$ K, the normalized SD is proportional to temperature, suggesting TA process[25] as a dominant mechanism responsible for the switching events. Above $T = 0.2$ K, the normalized SD decreases with temperature, which can be explained by PD process[26] or thermally-induced retrapping process[16].

The escape rate $\Gamma(I_C)$ can be obtained from the switching probability $P(I_C)$ via the relation[15] of $P(I_C) = [\Gamma(I_C)/(dI/dt)]\left\{1 - \int_0^{I_c} P(I')dI'\right\}$, where $dI/dt$ corresponds to the sweep rate and $\left\{1 - \int_0^{I_c} P(I')dI'\right\}$ is the probability of not switching until current $I_C$. Figure 2c shows $\Gamma(I_C)$ data (symbols) numerically obtained from $P(I_C)$ data at each temperature. $\Gamma(I_C)$ is almost temperature-independent in the MQT regime, while its slope decreases with temperature in the TA regime, but changes behavior in the PD regime.

The escape rate in the MQT regime[13] is given by $\Gamma_{MQT} = 12\omega_p(3\Delta U/h\omega_p)^{1/2}\exp[-7.2(1+0.87/Q)\Delta U/\hbar\omega_p]$, where $\Delta U = 2E_{J0}[(1-\gamma^2)^{1/2} - \gamma\cos^{-1}\gamma]$ is the barrier height of the tilted washboard potential, $\gamma = I/I_{C0}$ is the normalized current, $\omega_p = \omega_{p0}(1-\gamma^2)^{1/4}$ is Josephson plasma frequency, $\omega_{p0} = (2eI_{C0}/\hbar C)^{1/2}$ is the plasma frequency in zero bias current, $C$ is the junction capacitance and $Q = 4I_C/\pi I_R$. The fitting results are depicted in Fig. 2c using solid lines. The best fit of $\Gamma_{MQT}$ at $T = 0.02$ K reveals that $I_{C0} = 192$ nA and $C = 70$ fF, respectively (see black solid line in Fig. 2c). It is noted that $C$ is comparable to $C_{eff} = 20$ fF previously estimated from $E_{Th}$.

The rate of thermal escape[15] is given by $\Gamma_{TA} = a_t(\omega_p/2\pi)\exp[-\Delta U/k_B T]$, where $a_t = (1+1/4Q^2)^{1/2}-1/2Q$ is a damping-dependent factor. Since the escape rate in the TA regime is dependent upon temperature, the normalized SD increases with temperature, as shown in Fig. 2b. Decreasing SD at higher temperatures, however, are observed in Fig. 2b, which is a typical behavior of the PD process[16] [17]. The phase particle escaped from the washboard potential well via the TA process can be repeatedly retrapped in the neighboring potential well as a result of strong dissipation and thermal fluctuations. Then the escape rate in the PD regime is given by $\Gamma_{PD} = \Gamma_{TA}(1-P_{RT})\ln(1-P_{RT})^{-1}/P_{RT}$, where $P_{RT}$ is the retrapping probability[16,27]. Here, $P_{RT}$ is obtained from integration of the retrapping rate $\Gamma_{RT} = \omega_{p0}[(I-I_{R0})/I_{C0}](E_{J0}/2\pi k_B T)^{1/2}\exp(-\Delta U_{RT}/k_B T)$, where $I_{R0}$ is the fluctuation-free retrapping current, $\Delta U_{RT} = (E_{J0}Q_0^2/2)[(I-I_{R0})/I_{C0}]^2$ is a retrapping barrier, and $Q_0 = 4I_{C0}/\pi I_{R0}$ is the fluctuation-free quality factor. The fitting results of $\Gamma_{TA}$ and $\Gamma_{PD}$ are shown in Fig. 2c.

When we fit the $P(I_C)$ data using the TA and PD model, we obtain the escape temperature $T_{esc}$, which is perceived by the escaping phase particle. Figure 2d shows $T_{esc}$, obtained from the TA (black symbos) and PD (red) models, respectively. At temperature below the crossover temperature $T^*_{MQT}$ between the MQT and TA regimes, $T_{esc}$ is almost constant to be ~ 0.1 K, while it increases linearly with $T$ in the TA regime between $T^*_{MQT}$ and $T^*_{TA}$. Above $T^*_{TA}$ ~ 0.2 K, the PD model is more appropriate than the TA model for determining $T_{esc}$. The overall coincidence between $T_{esc}$ and $T$ supports the validity of our fitting results for the switching probability and the escape rate.

The crossover temperatures of $T^*_{MQT}$[25] and $T^*_{TA}$[28] are given by $T^*_{MQT} = a_t\hbar\omega_p/2\pi k_B$ and $T^*_{TA} \sim E_{J0}[1-(4/\pi Q)]^{3/2}/30 k_B$, respectively. Since $T^*_{MQT}$ is proportional to $I_{C0}^{1/2}$ and the $I_C$ can be tuned by the application of gate voltage in the NW-based JJ, gate-induced adjustments of $T^*_{MQT}$ and other physical parameters responsible for the escaping process of the phase particle would be possible. Figure 3a shows change of the switching probability depending on the gate voltage. Application of $V_g = -30$ V results in a decrease in $I_C$, an increase in SD and a reduced slope of the escape rate in Fig. 3b. The TA model fit results in $T_{esc} = 0.13$ (0.17) K for $V_g = 0$ (-30) V, indicating that the effective temperature can be adjusted by $V_g$.

In conclusion, we have studied the escape process of the Josephson phase particle in

InAs NW JJ with varying temperature and gate voltage. The switching current distribution of the NW JJ at each temperature is understood by the MQT, TA and PD models, depending on temperature. Application of gate voltage to suppress the critical current results in an effective increase of the escape temperature. Our results indicate that the nano-hybrid JJ would be a good platform to develop gate-tunable superconducting phase qubits.

# Acknowledgements

This work was supported by a Korea University Grant.

# Figure Captions

**Figure 1.** (a) Temperature dependence of current-voltage (*I-V*) characteristics of device D1. Bias current was swept from negative to positive polarities. $I_C$ and $I_R$ indicate the switching and return currents, respectively. Inset: SEM image of a typical InAs NW Josephson junction. (b) Temperature dependence of $I_C$ (circles) and $I_R$ (triangles). Solid line is a theoretical fit (see text).

**Figure 2.** (a) Temperature dependence of the switching current distribution obtained from D1 with $V_g = 0$ V. Symbols mean the experimental data and solid lines are the theoretical fitting results, respectively. (b) Normalized standard deviations of the switching current distributions. Three distinct regimes of MQT, TA, and PD are indicated. (c) Temperature dependence of the escape rate (symbols) obtained from the switching current distributions in (a). Solid lines are the theoretical fitting results (see text). (d) The escape temperature followed by TA (circle) and PD (triangle) models, respectively. $T^*_{MQT}$ ($T^*_{TA}$) corresponds to the crossover temperature from MQT (TA) to TA (PD) regimes.

**Figure 3.** (a) The switching current distribution and (b) the escape rate obtained from device D2 with different gate voltages at $T = 25$ mK. Solid lines represent fitting results following the TA model.

# Reference

[1]     B. D. Josephson, Phys. Lett. **1**, 251 (1962).

[2]     J. Clarke and F. K. Wilhelm, Nature **453**, 1031 (2008).

[3]     H. Takayanagi and T. Kawakami, Phys. Rev. Lett. **54**, 2449 (1985).

[4]     Y.-J. Doh, J. A. van Dam, A. L. Roest, E. P. A. M. Bakkers, L. P. Kouwenhoven, and S. De Franceschi, Science **309**, 272 (2005).

[5]     P. Jarillo-Herrero, J. A. van Dam, and L. P. Kouwenhoven, Nature **439**, 953 (2006).

[6]     H. B. Heersche, P. Jarillo-Herrero, J. B. Oostinga, L. M. K. Vandersypan, and A. F. Morpurgo, Nature **446**, 56 (2007).

[7]     M. Veldhorst *et al.*, Nature Mater. **11**, 417 (2012).

[8]     V. Mourik, K. Zuo, S. M. Frolov, S. R. Plissard, E. P. A. M. Bakkers, and L. P. Kouwenhoven, Science **336**, 1003 (2012).

[9]     A. Das, Y. Ronen, Y. Most, Y. Oreg, M. Heiblum, and H. Shtrikman, Nature Phys. **8**, 887 (2012).

[10]    G. de Lange, B. van Heck, A. Bruno, D. J. van Woerkom, A. Geresdi, S. R. Plissard, E. P. A. M. Bakkers, A. R. Akhmerov, and L. DiCarlo, Phys. Rev. Lett. **115**, 127002 (2015).

[11]    T. W. Larsen, K. D. Petersson, F. Kuemmeth, T. S. Jespersen, P. Krogstrup, J. Nygård, and C. M. Marcus, Phys. Rev. Lett. **115**, 127001 (2015).

[12]    M. Tinkham, *Introduction to Superconductivity: Second Edition* (Dover Publications, 2004).

[13]    J. Clarke, A. N. Cleland, M. H. Devoret, D. Esteve, and J. M. Martinis, Science **239**, 992 (1988).

[14]    A. O. Caldeira and A. J. Leggett, Phys. Rev. Lett. **46**, 211 (1981).

[15]    T. A. Fulton and L. N. Dunkleberger, Phys. Rev. B **9**, 4760 (1974).

[16]    V. M. Krasnov, T. Bauch, S. Intiso, E. Hürfeld, T. Akazaki, H. Takayanagi, and P. Delsing, Phys. Rev. Lett. **95**, 157002 (2005).

[17]    J. Männik, S. Li, W. Qiu, W. Chen, V. Patel, S. Han, and J. E. Lukens, Phys. Rev. B **71**, 220509 (2005).

[18]    H. Courtois, M. Meschke, J. T. Peltonen, and J. P. Pekola, Phys. Rev. Lett. **101**, 067002 (2008).

[19]    D. Jeong, J.-H. Choi, G.-H. Lee, S. Jo, Y.-J. Doh, and H.-J. Lee, Phys. Rev. B **83**, 094503 (2011).

[20]    M. Jung *et al.*, ACS Nano **5**, 2271 (2011).

[21]    P. Dubos, H. Courtois, B. Pannetier, F. K. Wilhelm, A. D. Zaikin, and G. Schön, Phys. Rev. B **63**, 064502 (2001).

[22]    G.-H. Lee, D. Jeong, J.-H. Choi, Y.-J. Doh, and H.-J. Lee, Phys. Rev. Lett. **107**, 146605 (2011).

[23]    J.-H. Choi, G.-H. Lee, S. H. Park, D. C. Jeong, J.-O. Lee, H.-S. Sim, Y.-J. Doh, and H.-J. Lee, Nat. Commun. **4**, 2525 (2013).

[24]     J. M. Martinis, M. H. Devoret, and J. Clarke, Phys. Rev. B **35**, 4682 (1987).

[25]     H. Grabert, P. Olschowski, and U. Weiss, Phys. Rev. B **36**, 1931 (1987).

[26]     D. Vion, M. Götz, P. Joyez, D. Esteve, and M. H. Devoret, Phys. Rev. Lett. **77**, 3435 (1996).

[27]     V. M. Krasnov, T. Golod, T. Bauch, and P. Delsing, Phys. Rev. B **76**, 224517 (2007).

[28]     J. M. Kivioja, T. E. Nieminen, J. Claudon, O. Buisson, F. W. J. Hekking, and J. P. Pekola, Phys. Rev. Lett. **94**, 247002 (2005).

**Figure 1.**

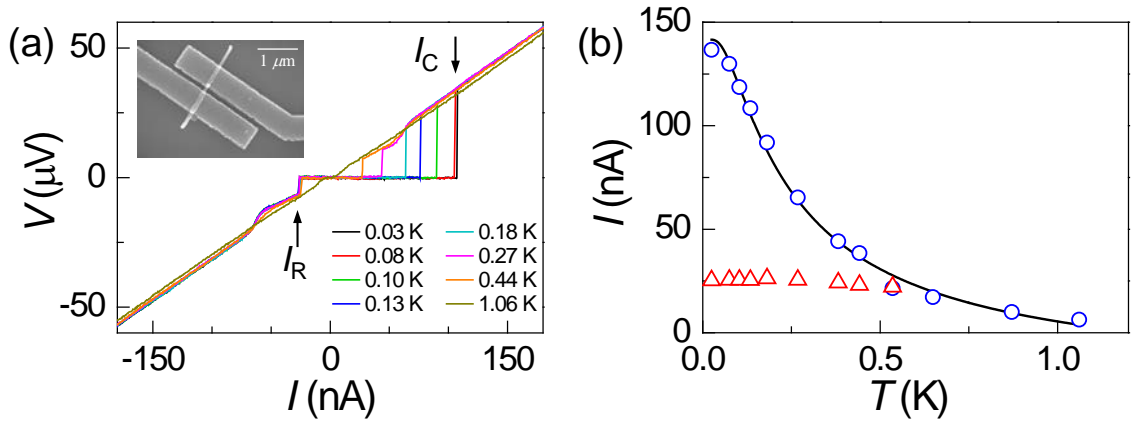

**Figure 2.**

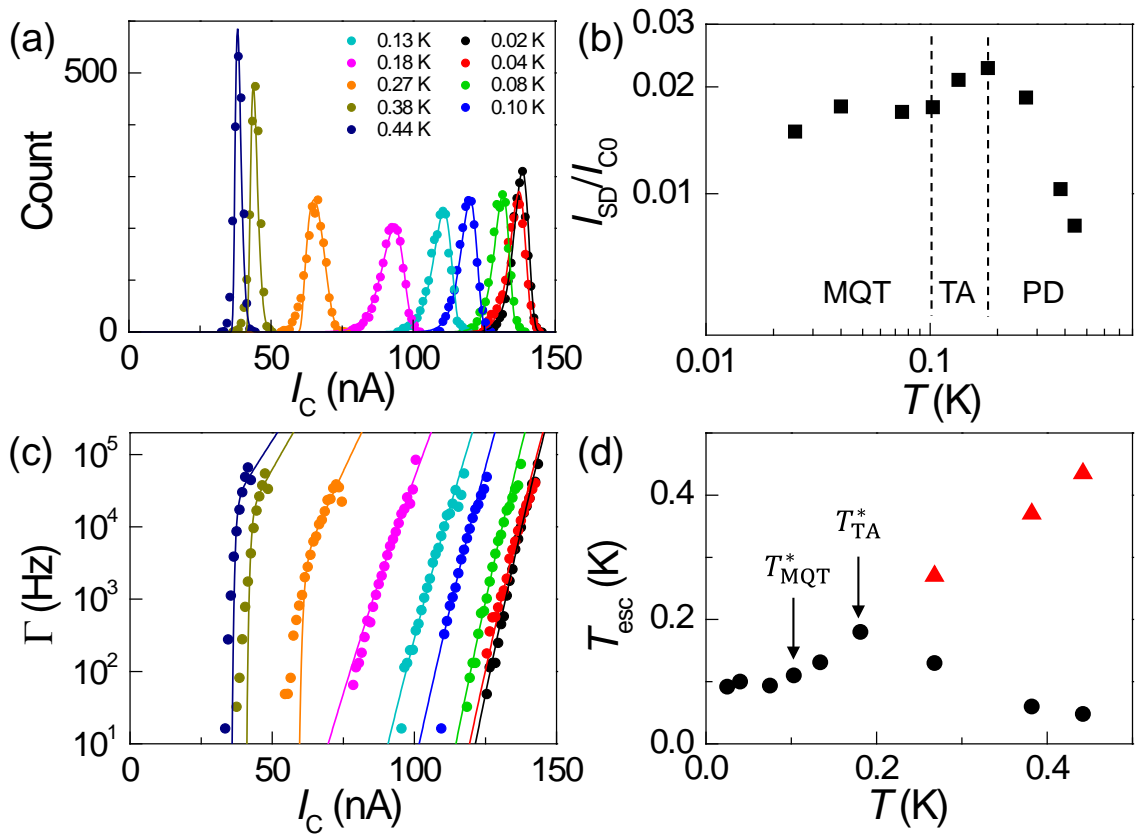

**Figure 3.**

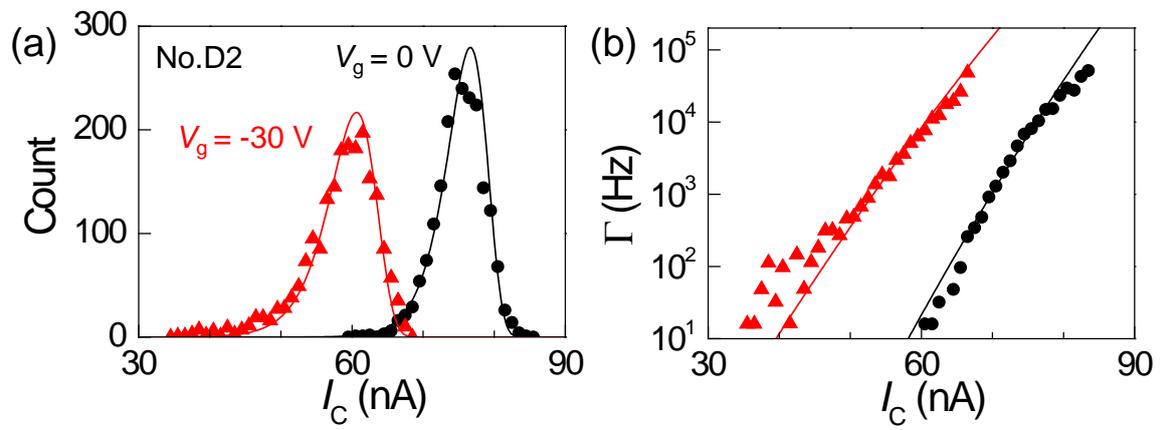